\documentclass{aa}  

\usepackage{graphicx}
\usepackage{txfonts}
\usepackage{lipsum}
\usepackage{subcaption}         
\usepackage{lscape}            
\usepackage{placeins}         
                                
\usepackage[colorlinks= true, linkcolor= magenta, citecolor = blue, urlcolor= cyan]{hyperref}

\begin{document}

   \title{Planet engulfment in the chemically anomalous HD 129171/HD 129209 pair}
 
   \author{Anne Rathsam\inst{1,2}
          \and
          Jorge Mel\'{e}ndez\inst{2,3}
          \and
          Rodolfo Smiljanic\inst{4}
          \and
          Fan Liu\inst{5,6}
          \and
          Lorenzo Spina\inst{3,7}
          }

   \institute{Leibniz-Institut f\"ur Astrophysik Potsdam (AIP), An der Sternwarte 16, 14482, Potsdam, Germany
   \and 
   Universidade de São Paulo, Instituto de Astronomia, Geofísica e Ciências Atmosféricas (IAG/USP), Departamento de Astronomia, Rua do Matão 1226, Cidade Universitária, 05508-090, SP, Brazil 
    \and
    INAF – Osservatorio Astronomico di Padova, Vicolo dell’Osservatorio 5, 35122 Padova, Italy
   \and
   Nicolaus Copernicus Astronomical Center, Polish Academy of Sciences, ul. Bartycka 18, 00-716 Warsaw, Poland
   \and
    National Astronomical Observatories, Chinese Academy of Sciences, Beijing 100101, China
    \and
    School of Physics and Astronomy, Monash University, Clayton 3800, Australia
    \and
    INAF - Osservatorio Astrofisico di Arcetri, Largo E. Fermi 5, 50125 Firenze, Italy \\
    \email{annerathsam@usp.br} 
    }

   \date{Received September 30, 2025}

  \abstract
   {Binary systems composed of stars with similar parameters should have identical chemical composition. However, many chemically anomalous pairs have been found in the literature, such as the binary HD 129171/HD 129209. It is still unclear whether these anomalies originate from inhomogeneities of protostellar clouds, with important implications for chemical tagging and theories of star formation, or if they are caused by a planet engulfment event suffered by one binary component.}
   {In this work, we measure precise differential abundances for the system HD 129171/HD 129209 to explore the planet engulfment hypothesis proposed in the literature. We focus particularly on the Be abundance, showing that this element can serve as a diagnostic of engulfment events for solar-type stars.} 
   {Atmospheric parameters were determined imposing spectroscopic equilibrium of iron lines. Masses and ages were estimated with the isochronal method. Li, Be, N and O abundances were determined via spectral synthesis. Other elemental abundances (up to Zn) were determined by equivalent width measurements. The spectra adopted in the analysis was gathered using UVES/ESO.}
   {We confirm the large difference in [Fe/H] (0.120 $\pm$ 0.004 dex) and A(Li) ($-$1.00 $\pm$ 0.02 dex) among the members of the pair, and the trend between differential abundances and condensation temperature of the elements. The binary system also shows detectable differences in Be abundances ($-$0.20 $\pm$ 0.04 dex). The abundance pattern of the pair is reasonably reproduced by an engulfment model of 11.2 M$_\oplus$ of rocky material.}
   {The difference in chemical abundances of the HD 129171/HD 129209 pair provides strong evidence in favor of the planet engulfment scenario. In this context, the detection of a Be difference among chemically inhomogeneous binary systems can be used as a diagnostic of rocky material ingestion suffered by a member of the pair.}

   \keywords{Stars: abundances -- Stars: atmospheres -- binaries: general -- Stars: solar-type}

   \maketitle
   \nolinenumbers

\section{Introduction}

It is widely accepted that stars in binary systems are formed from the collapse of the same molecular cloud and at around the same time. Thus, they share the same age and initial chemical composition. If the stars have similar parameters and therefore a similar internal structure, they should follow the same evolutionary path and be chemically identical.

However, recent works have shown the existence of binary pairs composed of ``twin" solar-type stars presenting significant chemical abundance variations, especially in refractory elements (e.g. \citealt{ramirez_binary, saffe_binary, nagar_binary, spina, jhon_binary, yong_binary, fan_nature}). The observed chemical anomalies allow for two possible interpretations: either the protostellar clouds are not homogeneous, a common assumption in star formation theories, or this is the result of a planet ingestion suffered by one member of the pair.

Both scenarios have important implications for theories of stellar and planetary system evolution. If protostellar clouds are not homogeneous, this challenges the idea that the chemical content of a star directly reflects the composition of the interstellar medium at the environment and epoch of its formation, which is the main premise of chemical tagging -- a method that aims to reconstruct the history of the Galaxy based on chemical abundances and kinematic information \citep{chem_tag, chem_tag1, chem_tag2}. On the other hand, if the inhomogeneity of the pairs is caused by planet engulfment episodes, this provides evidence that a non-negligible number of planetary systems evolve following highly dynamical routes, differing significantly from our Solar System, where the planets remained in approximately circular orbits. 

Many different dynamical processes have been suggested to lead a planet to migrate within its system, which may result in ejection from the system, collision between planets, or ingestion by the host star. In 3-body systems, perturbations from a highly inclined stellar or planetary companion can result in secular variations in a planet's eccentricity and inclination, a phenomenon known as von Zeipel-Lidov-Kozai mechanism \citep{zeipel, lidov, kozai, wu_murray, naoz2011, naoz2013, petrovich_kl, naoz_vzkl, church}. In systems with multiple widely separated eccentric or inclined planets, secular chaos -- the long-term dynamical evolution driven by mutual perturbations -- plays an important role \citep{wu, lithwick, teyssandier}. If the planets start in circular and coplanar orbits, planet–planet scattering during close encounters can significantly increase their eccentricities, eventually leading to ejections or collisions \citep{chambers, rasio_ford, wm, chatt, petrovich, church, marzari}. The existing torques due to interactions with the circumstellar disk can also drive migration \citep{trilling, rice_armitage, kley}. For a review on dynamics of planetary systems, we refer the reader to \citet{davies}, and for recent simulations of planet engulfment events, we refer to \citet{soares}, who found that although about half of the stars should ingest planets, a chemical signature could be detectable only in about 20\% of them.

Refractory elements offer an excellent way to distinguish between the planet engulfment or the proto-cloud inhomogeneity scenarios. Since they have high condensation temperatures ($\gtrsim$ 1000 K), they are the primary constituents of the rocky material in planetary systems -- terrestrial planets and cores of gaseous planets. In case of engulfment by a Sun-like star, this material is accreted by the star and then dissolved and mixed in the convective envelope, increasing the stellar surface abundances after the event  \citep{sandquist02}. This process, however, produces a metal-rich outer layer with an unstable mean molecular weight gradient, which triggers thermohaline mixing \citep{theado_vauclair, sevilla}. For fragile elements such as Li and Be, this thermohaline mixing induces depletion, as it can carry these elements below the convective zone, into their burning regions. Thus, the chemical enrichment caused by the engulfment disappears over time. Nevertheless, \citet{sevilla} demonstrated through simulations that the Li engulfment signature in stars with masses close to solar can be detected for $\geq$ 1 Gyr, adopting a conservative detection threshold of 0.05 dex.

Among the refractory elements, Li and Be are key to assess possible planetary engulfment scenarios. These elements are easily destroyed in the stellar interiors due to the low temperatures required for their burning, making their abundances highly sensitive to the size of the convective zone. Li is depleted mostly during the first Gyr of stellar evolution, especially during the pre-main sequence phase, and any enhancement of this element in one member of a binary twin pair could be an indication of chemical pollution occurred during the stellar lifetime \citep{sandquist02, sevilla}. Since Be is depleted at a higher temperature than Li ($\sim3.5\times10^6$ vs. $\sim2.5\times10^6$ K), and it does not show signs of depletion caused by stellar evolution effects in solar twins and analogs of higher mass \citep{tucci_maia, henrique}, its chemical signature can last longer. Furthermore, neither of these elements are produced by main sequence stars, and therefore the only way of enriching main sequence stars with Li and Be is via accretion of rocky material.

There is some indication that planet engulfment is the main cause of the observed inhomogeneities in binary systems. \citet{spina} studied 107 pairs with similar stellar parameters and found that the probability of finding a chemically anomalous pair increases with effective temperature (their Fig. 1). This is evidence in favor of the planet engulfment hypothesis, since lower-mass (colder) stars have deeper convective zones and are able to quickly dilute the rocky material. Additionally, the higher incidence of anomalous pairs among hotter stars cannot be explained by chemical inhomogeneities of the protostellar clouds. Their Fig. 2 shows that the polluting material is rich in refractory and poor in volatile elements, which is further evidence of the planet engulfment scenario, that predicts chemical differences trending with condensation temperature (e.g., \citealt{ramirez_binary, saffe_binary, oh18, nagar_binary, jhon_binary, behmard23, fan_nature}).

In this work, we analyze the system HD 129171/HD 129209, composed of two G-type stars. The pair presents an effective temperature difference of 23 K (5957 $\pm$ 9 vs. 5934 $\pm$ 9 K), a log $g$ difference of $-$0.09 dex (4.260 $\pm$ 0.024 vs. 4.350 $\pm$ 0.024 dex), a [Fe/H] difference of 0.11 dex (0.164 $\pm$ 0.008 vs. 0.049 $\pm$ 0.008 dex), and $\Delta$A(Li) = 1.03 dex \citep{jhon_binary}. Adopting spectra taken with the Robert G. Tull Coud\'e instrument at the Harlam J. Smith Telescope from the McDonald Observatory, with R$= \lambda/\Delta\lambda\sim$60 000 and SNR $\sim$350, \citet{jhon_binary} derived chemical abundances for 28 chemical elements, and found that the abundance difference pattern correlates with condensation temperature. \citet{jhon_binary} also demonstrated that the chemistry of the system can be explained by an engulfment of a 9.8$_{-1.6}^{+2.0}$ M$_\oplus$ rocky planet.

Using higher quality UVES/ESO data, we reanalyze the chemical pattern of HD 129171/HD 129209 to further explore the planet engulfment hypothesis. We are particularly interested in its beryllium (Be) abundance, as this element can help distinguish between the two possible scenarios -- a large Be anomaly would be consistent with a planet engulfment event, while a small Be difference could only be attributed to primordial differences due to inhomogeneities of the protostellar clouds. Thus, this element could potentially be used to identify stars that underwent planet engulfment events solely from the analysis of their spectra.

This paper is organized as follows: in Sect. \ref{sec:obs}, we describe the data acquisition and processing. In Sect. \ref{sec:param}, we present the determination of stellar parameters. Chemical abundances of the binary pair are explored in Sect. \ref{sec:chemistry}. In Sect. \ref{sec:engulf}, we model a planetary engulfment based on the abundance pattern of the system. Finally, our conclusions are summarized in Sect. \ref{sec:conc}.

\section{Observations and data reduction}
\label{sec:obs}
The binary pair spectra adopted in this work were observed with UVES/VLT at ESO Paranal, through our ESO program 113.26GJ.002 (PI: A. Rathsam). We used the dichroic mode, obtaining simultaneous UV and optical coverage in the standard setting of 346 nm + 580 nm, spanning 305.1 -- 387.0 nm and 478.8 -- 680.6 nm in the blue and red arms, respectively. The UV setup was used to determine Be, N, and O abundances, while the visible setup was used to derive the stellar atmospheric parameters and other elemental abundances. The resolving powers and SNRs achieved are R $\sim$60 000 and SNR $\sim$650 in the blue arm, and R $\sim$110 000 and SNR $\sim$800 in the red arm. 

Data processing was performed with the interface \texttt{IRAF}\footnote{Image Reduction and Analysis Facility, \url{https://iraf-community.github.io/}.}. Radial velocities were measured for each individual spectra using the task \texttt{rvidlines}, and the Doppler correction was performed with the task \texttt{dopcor}. We then combined separately blue and red observations for each star with the task \texttt{scombine}, and normalized with \texttt{continuum}. The normalization of the region around the Be doublet line at 3131 \r{A} was performed separately due to the high number of lines in the near-UV, which complicates the continuum placement, and will be described in Sect. \ref{sec:be}.

\section{Stellar parameters}
\label{sec:param}
The atmospheric parameters (effective temperature $T_\text{eff}$, log$g$, [Fe/H], and microturbulence velocity $v_t$) were determined imposing excitation and ionization equilibria of Fe I and II lines in a line-by-line differential analysis \citep{bedell2014, melendez2014}. Equivalent widths were measured for HD 129171, HD 129209, and the Sun using \texttt{IRAF}'s task \texttt{splot}, adopting the linelist from \citet{melendez2014}. The solar spectra was originally observed with UVES/VLT using the same setup described in Sect. \ref{sec:obs}, reflected on the surface of the asteroid Juno \citep{tucci_maia}. Spectroscopic equilibrium is performed with the code \texttt{q$^2$} (\textit{qoyllur-quipu}\footnote{\texttt{q$^2$} code: \url{https://github.com/astroChasqui/q2}.}, \citealt{q2}), which uses the stellar line analysis code \texttt{MOOG} (\citealt{moog}; \texttt{ABFIND} routine) and the Kurucz model atmospheres ATLAS9 \citep{kurucz}. Parameters were estimated for the binary pair adopting the Sun as the reference star. For validation, we also estimated atmospheric parameters for HD 129209 and the Sun using HD 129171 as reference. 

The results are shown in Table \ref{table:par}. We confirm that the binary pair is in fact composed of two very similar stars, with $\Delta T_\text{eff} =$ 24 $\pm$ 9 K and $\Delta$log $g = -$0.06 $\pm$ 0.02. We also confirm the chemical inhomogeneity of the system, that presents $\Delta$[Fe/H] $=$ 0.120 $\pm$ 0.004, consistent with the value of $\Delta$[Fe/H] $=$ 0.11 $\pm$ 0.01 reported by \citet{jhon_binary}. Additionally, the atmospheric parameters of HD 129209 are identical regardless of adopting the Sun or HD 129171 as the reference star, and the solar parameters are fully recovered when analyzed relative to the primary.

To find the projected rotational velocity ($v$ sin $i$) and the macroturbulence velocity ($v_\text{macro}$), we performed spectral synthesis of 4 Fe lines and 1 Ni line using \texttt{MOOG} (\texttt{synth} driver), as in \citet{leo}. These lines were selected by \citet{leo} due to the lack of contamination by nearby blending lines. Model atmospheres were interpolated from the Kurucz ATLAS9 grids and adopted for all subsequent syntheses. The final values of $v$ sin $i$ and $v_\text{macro}$, shown in Table \ref{table:par}, correspond to the mean of the individual measurements, with their standard deviations taken as uncertainties.

Ages and masses were estimated with \texttt{q$^2$} by generating probability distributions through the comparison of the spectroscopic atmospheric parameters with theoretical ones from Yonsei–Yale isochrones \citep{yi, kim}. The input [Fe/H] in \texttt{q$^2$} was corrected to account for the contribution of $\alpha$-elements to the overall metallicity, adopting Mg as a representative of $\alpha$-elements, as described in \citet{eu23}. The most probable values and their 1-$\sigma$ errors are reported in Table \ref{table:par}.

\begin{table*}[!ht]
\caption{Stellar parameters for the binary system HD 129171/HD 129209 and the Sun.} 
\label{table:par}    
\centering
\begin{tabular}{l c c c c c c c c}    
\hline\hline               
ID &  T$_\text{eff}$ & log $g$ & [Fe/H] & $v_t$ & $v$ sin $i$ & $v_\text{macro}$ & Age$^\text{c}$ & Mass$^\text{c}$ \\
& (K) & (dex) & (dex) & (km/s) & (km/s) & (km/s) & (Gyr) & (M$_\odot$) \\
\hline 
   HD 129171$^\text{a}$ & 5964 $\pm$ 7 & 4.280 $\pm$ 0.017 & 0.176 $\pm$ 0.005 & 1.22 $\pm$ 0.01 & 0.36 $\pm$ 0.31 & 4.78 $\pm$ 0.10 & 4.8$^{+0.6}_{-0.2}$ & 1.150$^{+0.014}_{-0.022}$ \\
   HD 129209$^\text{a}$ & 5940 $\pm$ 5 & 4.340 $\pm$ 0.014 & 0.056 $\pm$ 0.004 & 1.16 $\pm$ 0.01 & 0.10 $\pm$ 0.20 & 4.24 $\pm$ 0.09 & 5.1 $\pm$ 0.2 & 1.090$^{+0.017}_{-0.007}$ \\
   HD 129209$^\text{b}$ & 5940 $\pm$ 5 & 4.340 $\pm$ 0.015 & 0.056 $\pm$ 0.004 & 1.16 $\pm$ 0.01 & & & & \\
   Sun$^\text{b}$ & 5777 $\pm$ 6 & 4.440 $\pm$ 0.017 & 0.000 $\pm$ 0.006 & 1.00 $\pm$ 0.01 & & & &  \\
\hline 
\end{tabular}
\tablefoot{$^\text{a}$ Atmospheric parameters determined through the differential method relative to the Sun.\\ $^\text{b}$ Atmospheric parameters determined through the differential method relative to HD 129171. \\ $^\text{c}$ The age and mass of HD 129171 were estimated adopting the metallicity of its companion, as it likely represents its original (uncontaminated) metallicity.}
\end{table*}

\section{Chemical analysis}
\label{sec:chemistry}
\subsection{Li}

Li abundances for the binary pair were determined through spectral synthesis of the Li doublet lines near 6707.8 \r{A} using \texttt{MOOG}. We adopted the line list from \citet{melendez2012}, which takes blends and the hyperfine structure of the Li line into account. Since we expect solar-type stars to deplete all their $^6$Li during their pre-main sequence phase due to its extremely fragile nature \citep{sandquist02}, we considered only the contribution of $^7$Li to the total Li abundance. 3D NLTE corrections were calculated with the \texttt{Breidablik} code\footnote{\texttt{Breidablik} code: \url{https://github.com/ellawang44/Breidablik}.} ($-$0.096 dex for HD 129171 and $-$0.055 dex for HD 129209; \citealt{breidablik1, breidablik2}).

Observational errors (related to the quality of the data and depth of the absorption feature) were estimated by setting the minimum and maximum reasonable fits to the observed data and calculating the deviation from the best fit. Systematic errors (related to the uncertainties of the atmospheric parameters) were found by re-fitting the spectrum with a model atmosphere generated by adding/subtracting the uncertainty of each of the atmospheric parameters at a time while maintaining the others fixed. These errors are added in quadrature to obtain the systematic error. The total uncertainty is then $\sigma_{\text{A(Li)}} = \sqrt{\sigma_{\text{obs}}^2 + \sigma_{\text{sys}}^2}$, which is dominated by the observational error.

\begin{figure}[!ht]
    \centering
    \includegraphics[width=\linewidth]{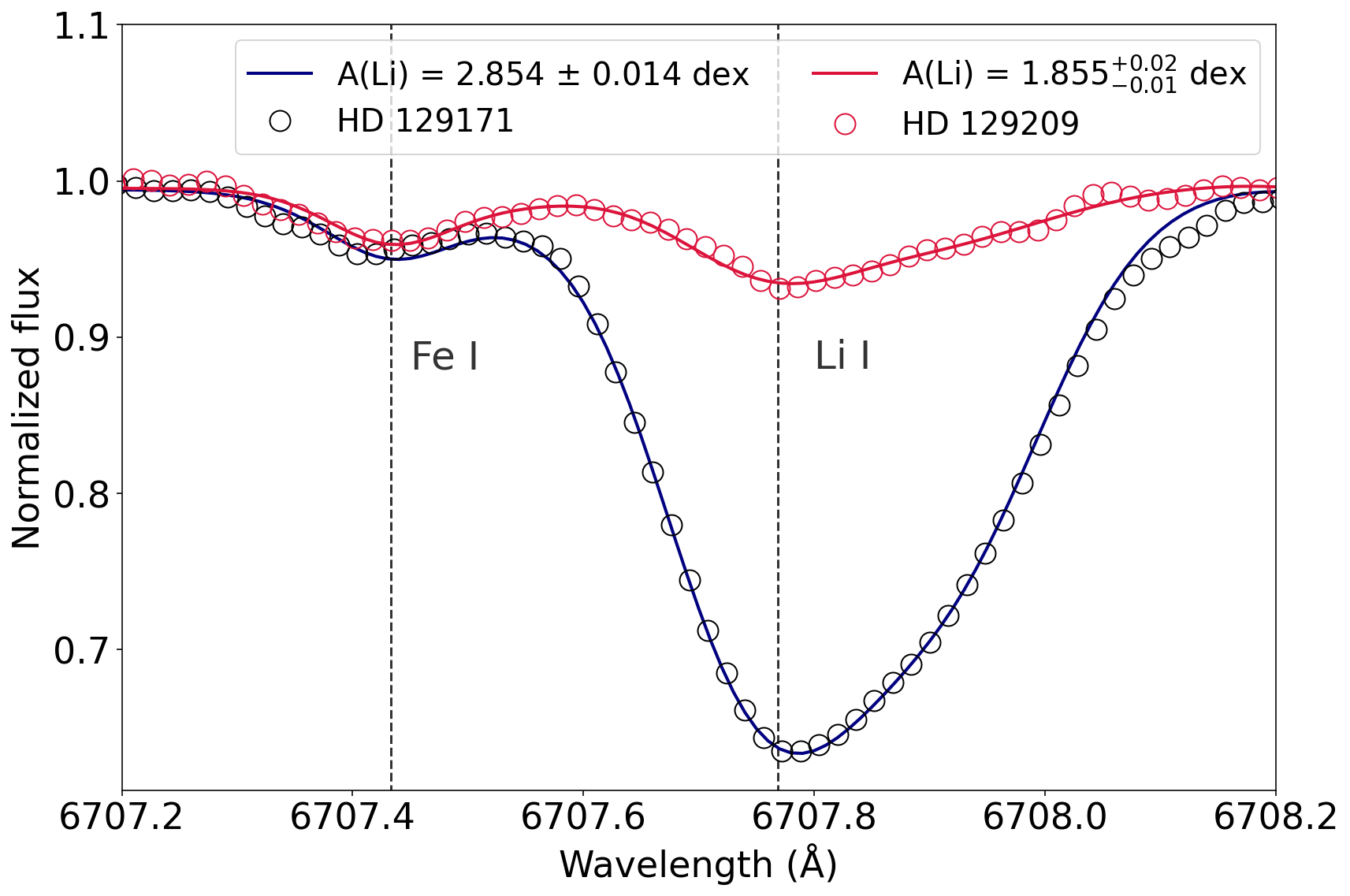}
    \caption{Spectral synthesis of the region of the $^7$Li resonance line for the pair HD 129171/HD 129209, with their corresponding 3D NLTE abundances as indicated (lines). Open circles represent observed spectra.}
    \label{fig:li_all}
\end{figure}

Figure \ref{fig:li_all} shows the comparison between the observed and the synthetic spectra of the binary pair. The difference in Li content between the stars is $\Delta$A(Li) $=$ 1.00 $\pm$ 0.02 dex, compatible with the (1D NLTE) value of 1.03 $\pm$ 0.09 dex reported by \citet{jhon_binary}. Since main sequence stars lack an internal Li production mechanism and the level of Li depletion due to stellar evolution should have affected both stars similarly (given their similar stellar parameters), this enrichment in the primary star can only be explained by accretion of rocky material -- either a terrestrial or a giant planet (which has a rocky core with an estimated mass in the order of $\sim$10 M$_\oplus$; e.g. \citealt{giant_planet1, giant_planet2, giant_planet3, giant_planet4}). The Li abundances of the pair are thus qualitatively consistent with a planet engulfment episode. 

\citet{c3po} developed a simple model to assess potential differences in the Li signature following the engulfment of either an Earth-like or a Jupiter-like planet, finding that both scenarios lead to indistinguishable Li enhancements. As discussed in \citet{c3po}, Earth-like material has a relatively low Li abundance, whereas in Jupiter-like material  the combination of low hydrogen content and high heavy-element abundances leads to a low Li mass fraction, despite potentially higher absolute Li content. Consequently, their work indicate that Li abundances alone are insufficient to distinguish between engulfment of a rocky planet or a rocky core of a giant planet.

From Figure \ref{fig:li_all}, we see that the right wing of the Li line is not well reproduced for HD 129171, which could indicate a non-negligible $^6$Li content. If confirmed, this would be further evidence supporting the engulfment scenario -- as mentioned before, $^6$Li should have been completely (or almost completely) depleted during the pre-main sequence phase. Indeed, \citet{li_isotopes} modeled planet ingestion by main sequence solar-type stars which predicted an abundance increase in both Li isotopes following such events. A $^6$Li signature has been sought in stars with planets, but mostly with null results (e.g., \citealt{reddy, mandell, ghezzi}).  

Nevertheless, before interpreting this feature as a genuine isotopic signature, it is important to note that spectral lines in solar-type stars show asymmetries caused by convective motions in their atmospheres \citep{gray}, which 1D LTE analyses cannot reproduce. These asymmetries strengthen the red wing of absorption lines, producing an effect similar to that caused by $^6$Li absorption. Thus, a limited 1D LTE approach, such as presented here, may result in a spurious $^6$Li signal \citep{cayrel}. 

\begin{figure}[!ht]
    \centering
    \includegraphics[width=\linewidth]{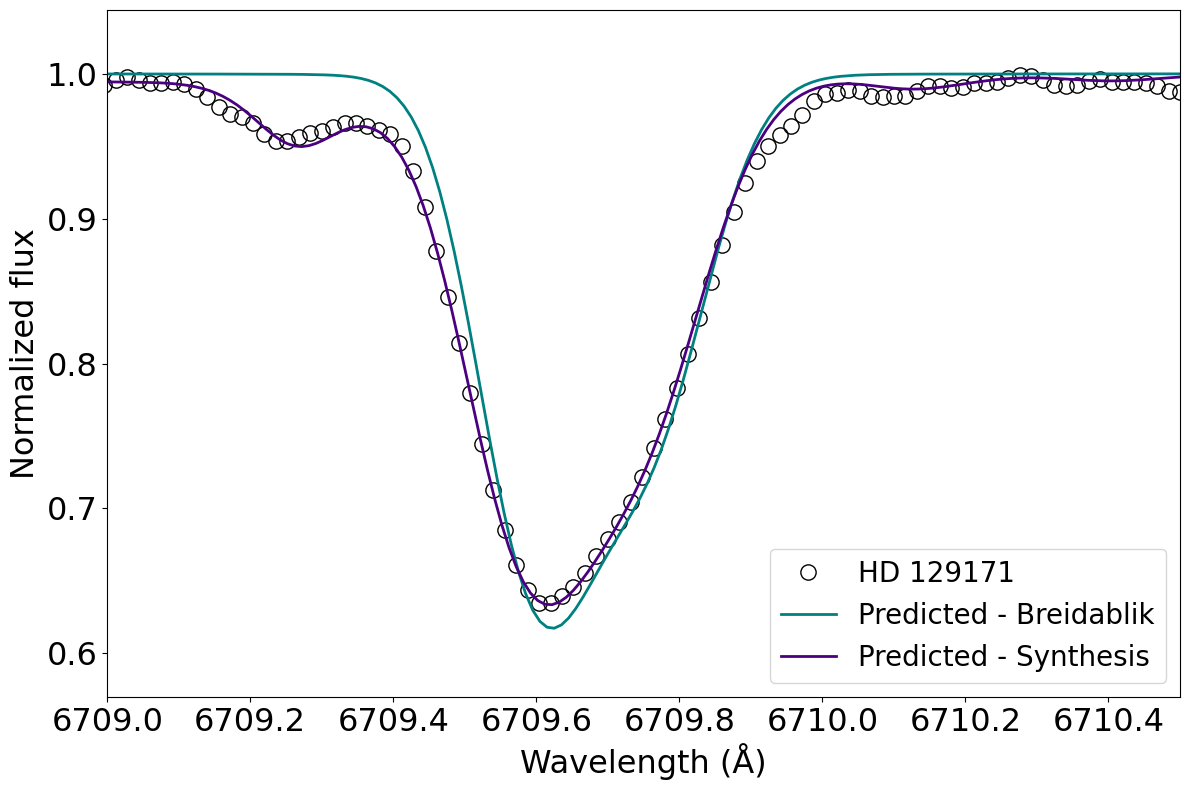}
    \caption{Region of the $^7$Li resonance line for HD 129171 (open circles), over-plotted with the predicted spectrum from \texttt{Breidablik} and the synthetic spectrum generated with \texttt{MOOG}. Note that the \texttt{Breidablik} profile considers only $^7$Li, while the \texttt{MOOG} profile also includes nearby blends. The observed and synthetic spectrum from \texttt{MOOG} were shifted to match the central line wavelength from \texttt{Breidablik}.}
    \label{fig:li_breidablik}
\end{figure}

To check the effect of a more refined spectral synthesis on HD 129171's Li line, we adopted the \texttt{Breidablik} code to interpolate its 3D NLTE $^7$Li line profile, shown in Figure \ref{fig:li_breidablik}. As can be seen, the line profile is still not well reproduced by the synthetic spectrum. Part of the discrepancy can be explained by nearby blends to the Li line (not considered by \texttt{Breidablik}), especially at the left wing, that is strongly affected by C and N abundances. However, the possibility of a non-zero $^6$Li content could not be discarded. Further investigations could be conducted adopting 3D model atmospheres, NLTE spectral syntheses and considering blending species.

\subsection{Be}
\label{sec:be}

As mentioned in Sect. \ref{sec:obs}, the near-UV spectral region, which includes the Be doublet lines at $\sim$3131 \r{A}, contains multiple atomic and molecular transitions, which complicates the normalization process, since there is no visible continuum reference. This region must then be treated carefully. 

To perform the normalization, we created a first normalized spectrum for each star with the \texttt{continuum} task on \texttt{IRAF} and generated a synthetic spectrum with \texttt{MOOG} adjusted to provide a reasonable fit to this observed normalized spectrum. Then, 12 additional observed spectra were created for each star by applying small differences to the inclination of the first normalized spectrum. These new spectra were compared internally, against the synthetic spectrum, and against the normalized solar spectrum from \citet{tucci_maia}, which is anchored to the superb continuum normalization of the \citet{kurucz84} solar spectrum. The final normalized spectra were chosen as those providing the best match to the synthetic spectrum, ensuring a consistent continuum shape across all stars. 

\begin{figure}[!ht]
    \centering
    \includegraphics[width=\linewidth]{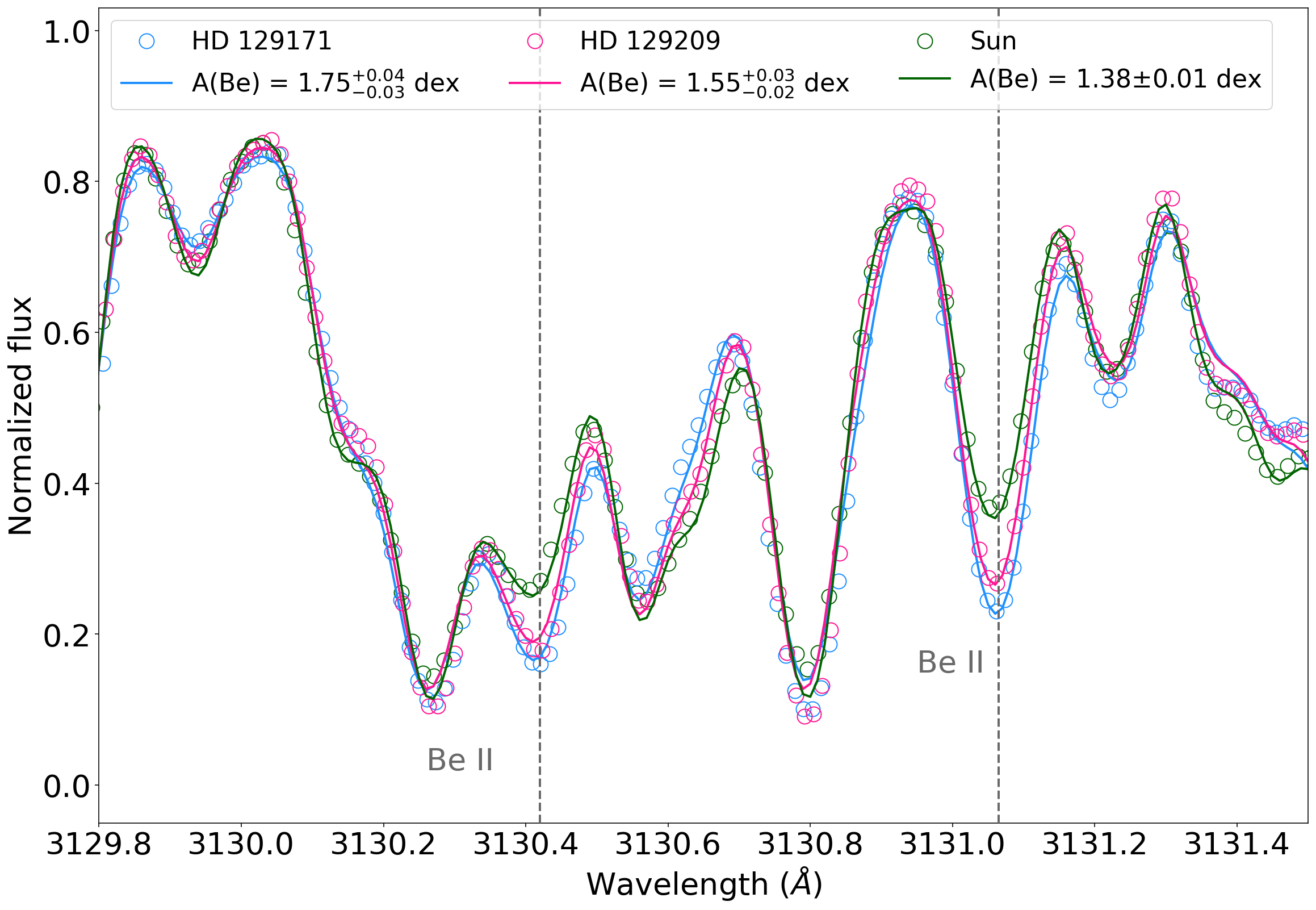}
    \caption{Spectral synthesis of the region of the $^9$Be resonance line for the pair HD 129171/HD 129209 and the Sun, with their corresponding LTE abundances as indicated (lines). Open circles represent observed spectra.}
    \label{fig:be}
\end{figure}

The spectral synthesis was then carried with the \texttt{synth} driver in \texttt{MOOG}, adopting the line list from \citet{tucci_maia}. Since systematic errors are too small compared to errors on the continuum placement, line depth, and due to the numerous blends, only observational errors were estimated. Figure \ref{fig:be} shows the comparison between observed and synthetic spectra near the Be doublet region for HD 129171, HD 129209, and the Sun, adopting A(Be)$_\odot =$ 1.38 $\pm$ 0.01 dex \citep{asplund09, tucci_maia}. The abundance difference among the binary pair is $\Delta$A(Be) $=$ 0.20 $\pm$ 0.04, indicating that HD 129171 is $\sim$1.6 times richer in Be than its companion. Since Be differences are not expected to emerge from a possible inhomogeneity of this system's birth cloud, the detected $\Delta$A(Be) among the pair is an additional evidence of a planet engulfment suffered by the primary star.

\subsection{N and O}

\begin{figure}[!ht]
    \centering
    \includegraphics[width=\linewidth]{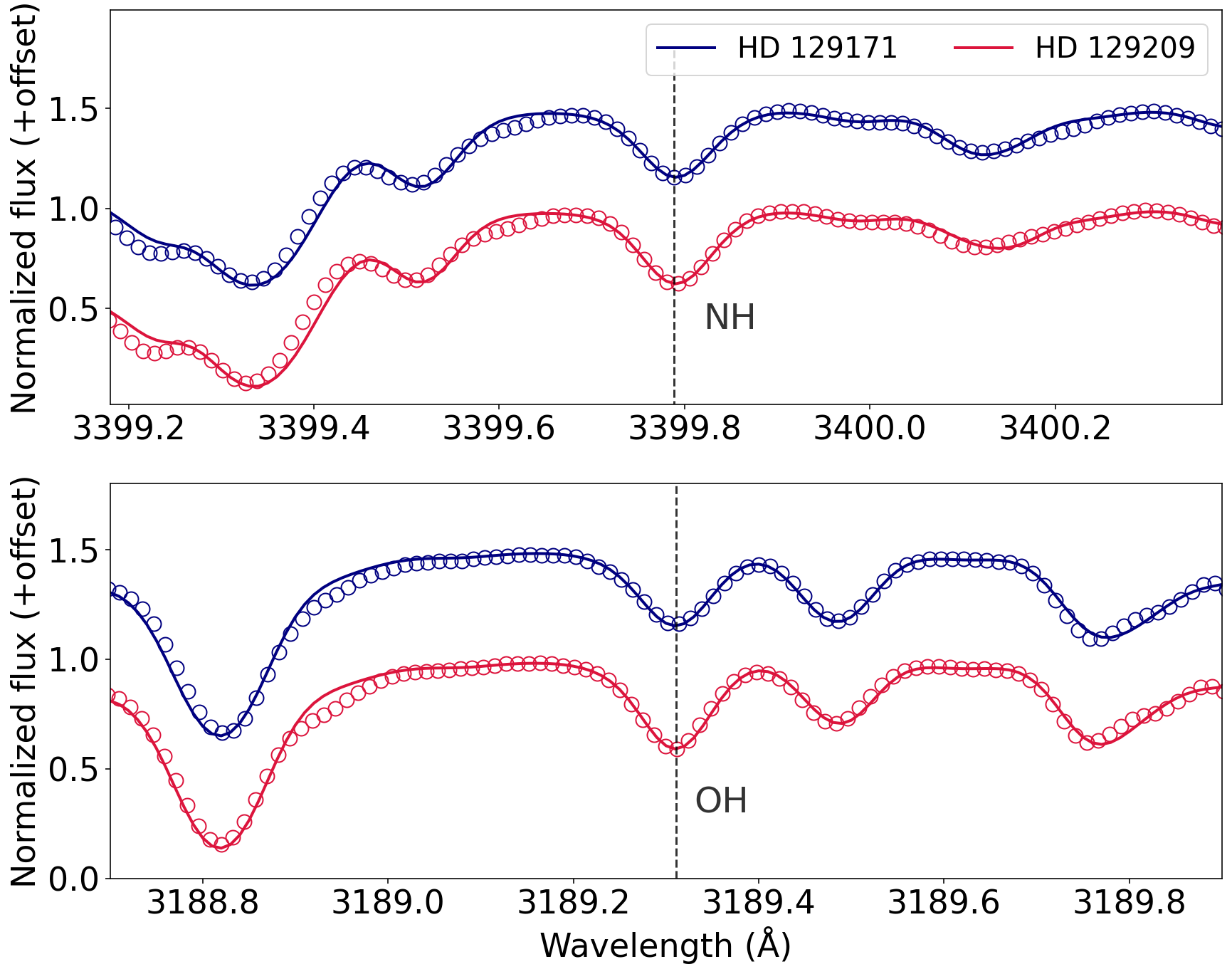}
    \caption{Spectral synthesis of the regions around the NH line (top) and OH 3189.3 \r{A} line (bottom) for HD 129171 and HD 129209 (lines). Open circles represent observed spectra. An offset was applied to the normalized flux of HD 129171 for better visualization.}
    \label{fig:nh_oh}
\end{figure}

To better determine the abundance pattern of the HD 129171/HD 129209 pair, measurements of low condensation temperature elements (volatiles), such as N and O, are essential. However, on UVES spectra, all detectable lines (from NH and OH molecules) fall on the far blue side of the spectrum\footnote{The [O I] line at $\sim$ 6300 \r{A} was visible in our data, but it was too shallow to allow a precise measurement.}, with many near contaminant species, preventing measurements of equivalent widths. Therefore, to accurately determine N and O abundances, we performed spectral synthesis using the observed molecular features at 3399.8 \r{A} (NH), 3189.3 \r{A}, 3218.1 \r{A}, and 3255.5 \r{A} (OH) with \texttt{MOOG}. Line lists were generated with the code \texttt{linemake}\footnote{\texttt{linemake} code: \url{https://github.com/vmplacco/linemake}.} \citep{linemake}. Errors for A(N) were estimated as observational errors (like for Be), while the uncertainty in the O abundance was taken as the standard deviation of individual measurements. The resulting abundances are shown in Table \ref{table:abundances}. The NH fit and an example of an OH fit are shown in Figure \ref{fig:nh_oh}.

\subsection{Elements up to \texorpdfstring{$Z=30$}{Z=30}}

Aside from Li, Be, N, and O, elemental abundances were determined through equivalent width measurements using the task \texttt{splot} on \texttt{IRAF}. Abundances were found with the code \texttt{q$^2$}, adopting a line-by-line differential analysis relative to the Sun and to HD 129171. The line list was taken from \citeauthor{melendez2014} (\citeyear{melendez2014}, same list adopted for the Fe measurements).

\begin{figure}[!ht]
    \centering
    \includegraphics[width=\linewidth]{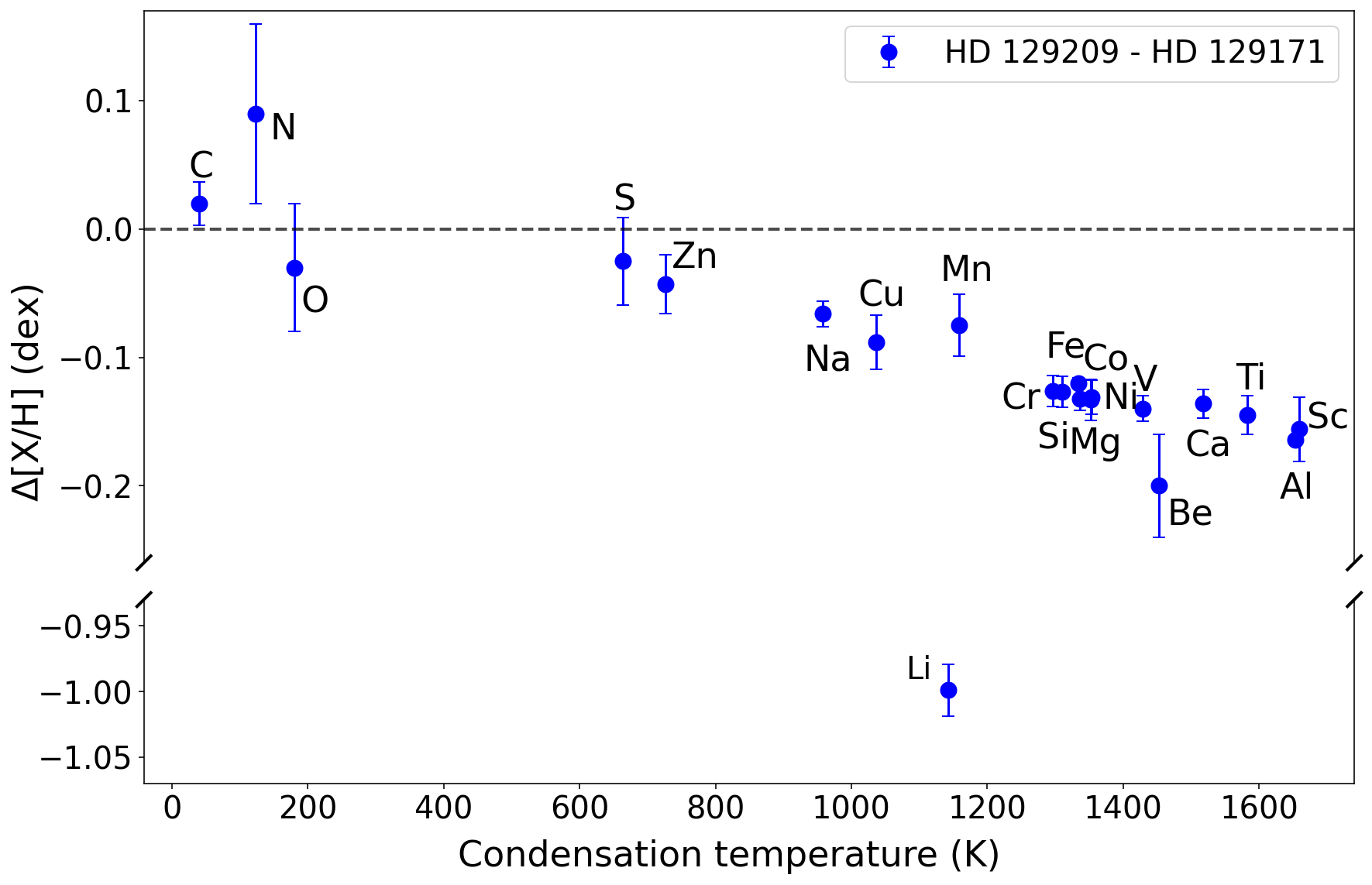}
    \caption{Differential abundances of HD 129209 relative to HD 129171 versus condensation temperature of the elements (from \citealt{lodders2003}).}
    \label{fig:ab_tcond}
\end{figure}

Differential abundances of the system are presented in Table \ref{table:abundances}, while Figure \ref{fig:ab_tcond} shows the abundances of HD 129209 relative to HD 129171 as a function of the condensation temperature of the elements (from \citealt{lodders2003}). The abundance pattern of the binary pair is consistent with a planet engulfment episode -- if HD 129171 ingested a refractory-rich planet in its past, HD 129209 would appear depleted in refractories relative to the primary, which is exactly what is seen in Figure \ref{fig:ab_tcond}.

\begin{table*}
\caption{Differential abundances of the system HD 129171/HD 129209.}  
\label{table:abundances}    
\centering
\begin{tabular}{l c c c c c c}    
\hline\hline               
Z & Element & Number & $T_\text{cond}^\text{a}$ & HD 129171 - Sun & HD 129209 - Sun & HD 129209 - HD 129171 \\
 & & of lines & (K) & $\Delta$[X/H] (dex) & $\Delta$[X/H] (dex) & $\Delta$[X/H] (dex) \\
\hline 
3 & Li$^\text{b}$ & 1 & 1142 & 2.854 $\pm$ 0.014 & 1.855$^{+0.020}_{-0.010}$ & $-$0.999 $\pm$ 0.021 \\
4 & Be$^\text{c}$ & 1 & 1452 & 1.75$^{+0.04}_{-0.03}$ & 1.55$^{+0.03}_{-0.02}$ & $-$0.20 $\pm$ 0.04 \\
6 & C & 3 & 40 & 0.027 $\pm$ 0.020 & 0.047 $\pm$ 0.008 & 0.020 $\pm$ 0.017 \\
7 & N$^\text{c}$ & 1 & 123 & 7.31 $\pm$ 0.05 & 7.4 $\pm$ 0.05 & 0.09 $\pm$ 0.07 \\
8 & O$^\text{c}$ & 3 & 180 & 8.55 $\pm$ 0.03 & 8.52 $\pm$ 0.03 & $-$0.03 $\pm$ 0.05 \\
11 & Na & 3 & 958 & 0.131 $\pm$ 0.017 & 0.065 $\pm$ 0.006 & $-$0.066 $\pm$ 0.011 \\
12 & Mg & 3 & 1336 & 0.182 $\pm$ 0.004 & 0.05 $\pm$ 0.011 & $-$0.132 $\pm$ 0.012 \\
13 & Al & 2 & 1653 & 0.21 $\pm$ 0.04 & 0.04 $\pm$ 0.04 & $-$0.164 $\pm$ 0.002 \\
14 & Si & 13 & 1310 & 0.193 $\pm$ 0.018 & 0.065 $\pm$ 0.015 & $-$0.127 $\pm$ 0.012 \\
16 & S & 4 & 664 & 0.07 $\pm$ 0.09 & 0.04 $\pm$ 0.11 & $-$0.03 $\pm$ 0.03 \\
20 & Ca & 10 & 1517 & 0.192 $\pm$ 0.011 & 0.056 $\pm$ 0.010 & $-$0.136 $\pm$ 0.011 \\
21 & Sc & 9 & 1659 & 0.24 $\pm$ 0.05 & 0.08 $\pm$ 0.03 & $-$0.156 $\pm$ 0.025 \\
22 & Ti & 21 & 1582 & 0.199 $\pm$ 0.026 & 0.054 $\pm$ 0.017 & $-$0.145 $\pm$ 0.015 \\
23 & V & 9 & 1429 & 0.189 $\pm$ 0.017 & 0.049 $\pm$ 0.017 & $-$0.140 $\pm$ 0.010 \\
24 & Cr & 16 & 1296 & 0.165 $\pm$ 0.020 & 0.040 $\pm$ 0.020 & $-$0.126 $\pm$ 0.012 \\
25 & Mn & 5 & 1158 & 0.12 $\pm$ 0.04 & 0.040 $\pm$ 0.009 & $-$0.075 $\pm$ 0.024 \\
26 & Fe & 99 & 1334 & 0.176 $\pm$ 0.005 & 0.056 $\pm$ 0.004 & $-$0.120 $\pm$ 0.004 \\
27 & Co & 9 & 1352 & 0.193 $\pm$ 0.027 & 0.061 $\pm$ 0.017 & $-$0.133 $\pm$ 0.016 \\
28 & Ni & 17 & 1353 & 0.185 $\pm$ 0.016 & 0.054 $\pm$ 0.012 & $-$0.131 $\pm$ 0.013 \\
29 & Cu & 3 & 1037 & 0.158 $\pm$ 0.023 & 0.070 $\pm$ 0.007 & $-$0.088 $\pm$ 0.021 \\
30 & Zn & 2 & 726 & 0.10 $\pm$ 0.03 & 0.054 $\pm$ 0.003 & $-$0.043 $\pm$ 0.023 \\
\hline 
\end{tabular}
\tablefoot{$^\text{a}$ From \citet{lodders2003}. \\ $^\text{b}$ 3D NLTE A(Li) $= \log (\text{N}_\text{Li}/\text{N}_\text{H}) + 12$.\\ $^\text{c}$ 1D LTE A(X) $= \log (\text{N}_\text{X}/\text{N}_\text{H}) + 12$.}
\end{table*}

Another phenomenon that can cause abundance differences in binary pairs, independent of formation and accretion of planets, is atomic diffusion (e.g., \citealt{michaud}), that consists in a global movement of particles due to gradients (of pressure, temperature, and/or concentration) and forces acting in the stellar interior. \citet{fanliu_diffusion}, for example, studied seven binary pairs and found that differences in [Fe/H] and chemical composition among the system members could be qualitatively explained by atomic diffusion effects for pairs with $\Delta$log $g$ $>$ 0.05 dex (similar to the derived $\Delta$log $g$ for HD 129171/HD 129209, $-$0.06 dex). However, both the observed and predicted abundance differences from models including diffusion are on the order of $\sim$0.02 dex, which is much smaller than the typical $\Delta$[X/H] of HD 129171/HD 129209 (average |$\Delta$[X/H]| $=$ 0.15 dex; median |$\Delta$[X/H]| $=$ 0.13 dex). Additionally, chemical inhomogeneities arising from atomic diffusion are not expected to correlate with condensation temperature. Thus, even if part of the chemical differences of the pair are due to atomic diffusion effects, this is still unable to fully explain the observed chemical pattern of the system. 

\section{Planet engulfment}
\label{sec:engulf}

To further explore the planet ingestion hypothesis proposed by \citet{jhon_binary}, we estimated the engulfed planetary mass required to reproduce the abundance pattern seen in Figure \ref{fig:ab_tcond} using the \texttt{Terra} code\footnote{\texttt{Terra} code: \url{https://github.com/ramstojh/terra}.} \citep{jhon_terra}. The code assumes that the engulfment occurs after the star has reached the main sequence and that internal mixing is negligible; therefore, the inferred engulfed mass should be regarded as a lower limit. Another important assumption is that the hypothetical ingested planet closely resembles those in the Solar System: \texttt{Terra} adopts solar abundances from \citet{asplund21} as the baseline for the stellar system (with a correction considering the stellar [Fe/H]) and models engulfment of rocky material using a combination of terrestrial abundances from \citet{earth} and chondrite abundances from \citet{chondrites}. However, \citet{planets_composition} showed that rocky exoplanets around solar-type stars can, in principle, be chemically different than Solar System planets. Even so, this approach is still informative, especially given their finding that both the Sun and Earth lie close to the medians of the predicted composition distributions.

Since the birth environment of HD 129171/HD 129209 may have contained a different Li content than the solar protocloud, we adopted the Li abundance of HD 129209 as the base A(Li), which is expected to represent the initial abundance of the system (depleted by effects of stellar evolution, which would have affected both stars similarly). We tested different chondrite abundances available in the code (CM, CI, H, L, and LL from \citealt{chondrites}) and updated values from \citeauthor{lodders21} (\citeyear{lodders21}, CM, CI, CV, H, L, LL, and EH), and found that the CV chondrite abundances from \citet{lodders21} provided the best match to the observed data. Additionally, we adopted as solar A(Be) the revised value from \citeauthor{amarsi} (\citeyear{amarsi}, 1.21 $\pm$ 0.05 dex), which also improved the agreement between the model and the data.

\begin{figure}[!ht]
    \centering
    \includegraphics[width=\linewidth]{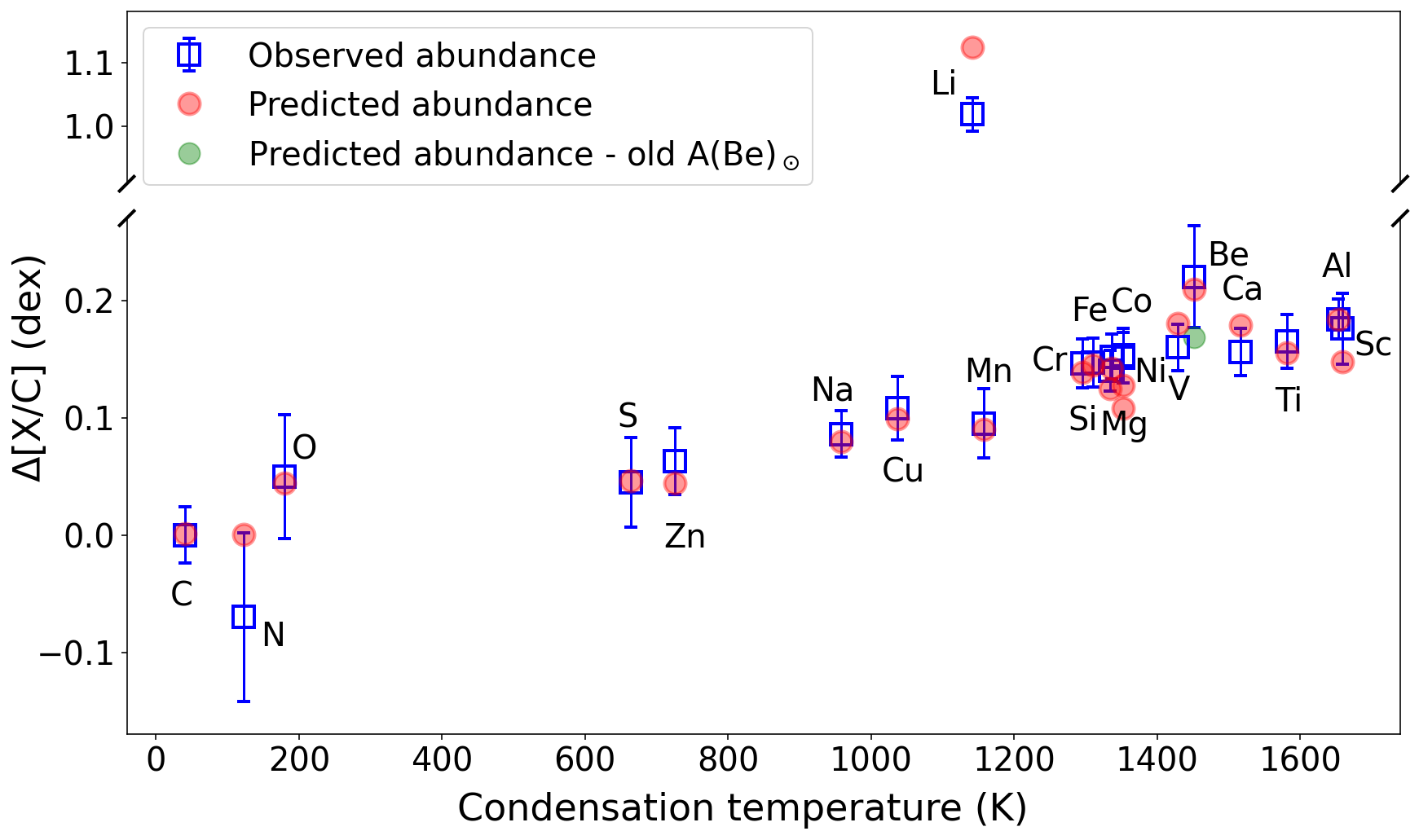}
    \caption{Comparison of the observed differential abundances of HD 129209 relative to HD 129171 (open squares) and the predicted abundances estimated from the model of an engulfment of 11.2 M$_\oplus$ of rocky material (filled circles). Our adopted solar A(Be) is 1.21 dex \citep{amarsi}, but we also indicate the predicted beryllium abundance adopting the older 1.38 dex estimate from \citeauthor{asplund21} (\citeyear{asplund21}, green circle).}
    \label{fig:engulfment}
\end{figure}

The estimated planetary mass ingested by HD 129209 is 11.2 M$_\oplus$ (0.1 M$_\oplus$ of Earth-like composition and 11.1 M$_\oplus$ of CV chondrite-like material, \citealt{lodders21}). The comparison between the observed differential abundances of the system and the predicted differential abundances from the engulfment model are presented in Figure \ref{fig:engulfment}, which shows a reasonable match -- the predicted values agree with the observed ones within 1-$\sigma$ for almost all elements. The predicted Be abundance difference adopting the solar A(Be) from \citeauthor{asplund21} (\citeyear{asplund21}, 1.38 $\pm$ 0.09 dex) as the base A(Be) is also indicated.

The amount of engulfed rocky material could, in principle, have originated from a single 11.2 M$_\oplus$ planet or from several small planets whose masses add up to 11.2 M$_\oplus$. As \citet{lane} demonstrated, it is possible to distinguish between these scenarios when the stellar convective zone is relatively thin -- a large object would not be fully destroyed in the convective envelope, and thus the observed pollution would only reflect the composition of the planet's outer layers, whereas in the case of engulfment of several small bodies, the planets would be completely disrupted before crossing the boundary with the radiative interior, and their full bulk composition would be mixed into the stellar envelope. However, for stars with masses close to solar, as the system studied here, the convective envelope is sufficiently deep that engulfed material -- whether delivered by a single large body or by multiple smaller ones -- is expected to be fully disrupted and homogenized before reaching the radiative interior (\citealt{lane}, their Table 4). In this regime, the resulting chemical signature does not retain information about the delivery mechanism, and our results therefore do not allow us to distinguish between these two scenarios. Additionally, even if the core of a large body survives until it reaches the stellar radiative region, the exact chemical signature would depend on on the internal structure of the planet(s), which is not known.

Figure \ref{fig:ab_tcond} shows that the revised lower A(Be)$_\odot$ from \citet{amarsi} (1.21 $\pm$ 0.05 dex) provides a better match to the observed differential abundance than the \citet{asplund21} value, which validates the \citet{amarsi} determination, reinforcing the difference between the solar photospheric and meteoritic (1.32 $\pm$ 0.03 dex, \citealt{asplund21}) Be abundance. Since the solar convective zone is unable to take Be into regions that are able to destroy it \citep{tucci_maia, henrique}, we do not expect Be depletion in the Sun due to its evolution. Some Be depletion may be expected in the solar photosphere relative to the primordial A(Be) in the protostellar cloud due to the formation of terrestrial planets, which gather refractory elements from its surroundings, leaving a refractory-depleted gas that is later accreted by the Sun as it evolves towards the pre-main sequence (as proposed by \citealt{melendez2009}). Alternatively, this depletion could be caused by the formation of giant planets, which creates pressure traps that prevent dust and pebbles from the protoplanetary disc from falling onto the host star \citep{booth_owen, huhn_bitsch}. Additionally, part of this difference might be due to Be production in the early Solar System via cosmic ray spallation \citep{dwek}. 

\section{Discussion and conclusions}
\label{sec:conc}

In this work, we determined atmospheric parameters, masses, ages, and measured precise chemical abundances for elements up to $Z=30$ for the binary pair HD 129171/HD 129209 adopting UVES/ESO spectra. We confirmed the chemical inhomogeneity of the system and the trend between differential abundances and condensation temperature of the elements \citep{jhon_binary}, which is in qualitative agreement with the hypothesis that the primary star HD 129171 has ingested a rocky planet or planetary core at some point in its life, enriching its photosphere with refractory material.

The abundance pattern of the pair can be reasonably reproduced by an engulfment model generated with the \texttt{Terra} code \citep{jhon_terra} of 11.2 M$_\oplus$ (a combination of 0.1 M$_\oplus$ of terrestrial composition from \citealt{earth} and 11.1 M$_\oplus$ of CV chondrite-like material from \citealt{lodders21}, Figure \ref{fig:engulfment}), which represents a lower limit to the mass of the engulfed planet (or planets).

Assuming that the observed chemical differences are in fact due to planetary ingestion, we showed the importance of focusing on the light elements $^6$Li and Be, that are not produced in the stellar interior. Whereas the more common isotope $^7$Li is depleted throughout the main sequence phase of solar-type stars and the interpretation of its abundances are not straightforward (e.g., \citealt{eu23}), $^6$Li is completely destroyed during the pre-main sequence stage, and a detectable $^6$Li content can be interpreted as evidence of rocky material accretion \citep{li_isotopes}. However, due to the low temperatures required for its destruction ($\sim$2.2 $\times10^6$ K, easily reached by the convective envelopes of solar-type stars) and the expected thermohaline mixing following an engulfment event \citep{theado_vauclair, sevilla}, the $^6$Li signature of planet ingestion is expected to fade quickly. Albeit the HD 129171 Li line profile shows some asymmetry that could be due to $^6$Li, the $^6$Li/$^7$Li isotopic ratio is not accurately determined in a 1D analysis, since the asymmetry caused by the presence of $^6$Li on the 6707.8 \r{A} Li doublet is similar to the asymmetry due to convective motions \citep{cayrel, 6li_3d_nlte}, which can only be accurately described by 3D NLTE analyses, that are computationally expensive. A tentative 3D NLTE synthetic interpolated spectrum was generated with the \texttt{Breidablik} code, and the asymmetry still persists, which could be further indication of $^6$Li. The addition of nearby contaminant species to the Li doublet line would also help improve the match to the observed spectrum. 

As for Be, it is neither produced nor depleted by solar-type stars with masses larger or close to solar \citep{tucci_maia, henrique}, and local molecular cloud Be inhomogeneities are not expected. Thus, any detected A(Be) difference among similar solar-type binary stars can be used as a diagnostic of planet ingestion. Additionally, since its destruction requires larger temperatures than the Li isotopes, the Be signature is expected to last longer, making it easier to identify engulfment events via Be abundances.

\begin{acknowledgements}
AR thanks Fundação de Amparo à Pesquisa do Estado de São Paulo (FAPESP) for the Ph.D. funding (process no. 2023/07617-5). JM acknowledges the support from FAPESP (via process no. 2018/04055-8). Based on observations collected at the European Southern Observatory under ESO program 113.26GJ.002. 
\end{acknowledgements}

\bibliographystyle{aa} 
\bibliography{bibliography} 
\end{document}